\documentclass[conference]{IEEEtran}
\IEEEoverridecommandlockouts
\usepackage{cite}
\usepackage{amsmath,amssymb,amsfonts}
\usepackage{algorithmic}
\usepackage{graphicx}
\usepackage{textcomp}
\usepackage{xcolor}
\usepackage{hyperref}
\usepackage{fancybox}
\usepackage{ascmac} 
\usepackage{multirow} 
\usepackage{array}
\usepackage{caption}
\usepackage{subcaption}
\def\BibTeX{{\rm B\kern-.05em{\sc i\kern-.025em b}\kern-.08em
    T\kern-.1667em\lower.7ex\hbox{E}\kern-.125emX}}
\begin{document}

\title{
Do LLMs Make Neural Distinguishers Wise?
}

\author{\IEEEauthorblockN{Tatsuya Sakagami} 
\IEEEauthorblockA{\textit{Panasonic Holdings Corporation} \\
 1006 Kadoma, Kadoma City, Osaka\\
sakagami.tatsuya@jp.panasonic.com}
\and
\IEEEauthorblockN{Masashi Hisai}
\IEEEauthorblockA{\textit{Panasonic Holdings Corporation} \\
 1006 Kadoma, Kadoma City, Osaka\\
} 
\and
\IEEEauthorblockN{Naoto Yanai}
\IEEEauthorblockA{\textit{Panasonic Holdings Corporation} \\
 1006 Kadoma, Kadoma City, Osaka\\
yanai.naoto@jp.panasonic.com}
}

\maketitle

\begin{abstract}
Neural distinguishers are a cryptanalysis method for symmetric-key cryptography that trains machine learning models on pairs of plaintexts and ciphertexts with specific differences in order to recover a secret key.
To the best of our knowledge, no existing work has explored the use of large language models (LLMs) for neural distinguishers.
In this paper, we propose LLM-based neural distinguishers through a prompt design and conduct extensive experiments with them on SPECK-32/64 to investigate whether LLMs can strengthen neural distinguishers. 
We then found three key insights. 
First, by comparing the results of LLM-based neural distinguishers with ResNet in the existing work, we demonstrate that LLMs provide no observable improvement in the performance of neural distinguishers.
Second, we confirm that, at high rounds, the choice of differences is no longer effective for LLM-based neural distinguishers as well as ResNet. 
Third, we show that the performance of LLM-based neural distinguishers can be significantly improved by incorporating only the XOR operation results as a prompt design. 
\end{abstract}


\section{Introduction}\label{intro}

Machine learning has recently been employed in the field of cryptanalysis~\cite{Gohr,hospodar2011machine} beyond well-known applications. 
Neural distinguishers~\cite{Gohr} are an effective cryptanalysis method that leverages machine learning models trained on pairs of plaintexts and ciphertexts with specific differences to recover secret keys from unseen ciphertexts. 
Neural distinguishers also have the potential to improve traditional differential cryptanalysis~\cite{yadav2021differential-ml}. 
Various subsequent works of neural distinguishers have been conducted so far~\cite{baksi2022machine,hu2025enhancingneural, kimura2022outputprediction,lu2025enhanced,wang2024anewrelated-key,benamira2021deeper, hou2021improve,sun2022neural,bao2022enhancing,buacuieti2022deep}.

However, the accuracy of neural distinguishers in the above works is often low, primarily due to limitations in the underlying neural network architectures~\cite{Jiang}. 
It is a critical issue in practical cryptographic settings, such as in higher rounds of symmetric-key encryption.
Namely, exploring potential architectures of neural distinguishers is essential for cryptanalysis.  

In this paper, we focus on the use of \textit{large language models (LLMs)} in neural distinguishers. 
(We will refer to them as \textit{LLM-based neural distinguishers}.)
LLMs demonstrate strong performance in a wide range of data analysis tasks~\cite{Cheng} and are increasingly being adopted in cybersecurity research~\cite{Pedro, He,deng2024pentestgpt}. 

Consequently, the growing role of LLMs motivates their use for cryptanalysis.
However, it remains unclear whether the use of LLMs improves the accuracy of neural distinguishers.
This paper thus tackles the following research questions:

\begin{enumerate}
  \setlength{\itemsep}{0cm} 
  \renewcommand{\labelenumi}{\textbf{RQ\theenumi}} 
  \item Can LLMs improve performance of neural distinguishers? 
  \item How much can accuracy of LLM-based neural distinguishers be improved by enhancing the distribution of a training dataset?
  \item How do prompts have an impact on the performance of LLM-based neural distinguishers?
\end{enumerate}

The above research questions are \textit{non-trivial}.
In general, since factors that contribute to the inference of a machine learning model are opaque, it is difficult to directly reveal conclusions of the above question from existing works that do not employ LLMs. 
Even when using the same model, its performance can often be improved by enhancing the distribution of a training dataset~\cite{Jiang,seok2024novelapproach}. 
Therefore, we also examine the impact of enhancing the dataset. 
Furthermore, since the output of LLMs generally varies depending on the given instructions, known as \textit{prompts}, we also investigate the impact of prompts on LLM-based neural distinguishers.

To shed light on the above research questions, we propose LLM-based neural distinguishers and conduct extensive experiments to investigate whether LLMs can strengthen the threat posed by neural distinguishers.
In addition, by enhancing the distribution of a dataset and exploring prompts to the LLMs, we evaluate how these factors influence their performance.
When we utilize five kinds of LLM, including gpt‑oss\footnote{\url{https://openai.com/index/introducing-gpt-oss}} as an open‑weight model, and evaluate ciphertexts of SPECK~\cite{speck-2013} as our target encryption algorithm, we obtain three key insights as the answers to the research questions.

\begin{table*}[t]
    \centering
    \caption{Comparison of related works. 
    Checkmarks indicate which aspects each work investigates. 
    The checkmarks in the following table indicate that each scheme discusses the corresponding topic. 
    }
    \label{tab:related work}
    \begin{tabular}{c|c|c|c|c|c} \hline
         Reference & Use of LLMs & Neural distinguisher& Dataset distribution & Model & Block cipher  \\ \hline
         Wang et al. \cite{wang2024anewrelated-key} &  & \checkmark & \checkmark & CNN & SIMON, SPECK, SIMECK \\ \hline
         Benamira et al. \cite{benamira2021deeper} &  & \checkmark & \checkmark & ResNet & SIMON, SPECK\\ \hline
         Z.Hou et al. \cite{hou2021improve} &  & \checkmark & \checkmark & ResNet & SIMON, SPECK \\ \hline
         Bao et al. \cite{bao2022enhancing} &  & \checkmark &\checkmark & ResNet, DenseNet, SENet & SIMON, SPECK \\ \hline
         B{\u{a}}cuieți et al. \cite{buacuieti2022deep} &  & \checkmark &  & ResNet, Autoencoder & SPECK \\ \hline
         Jiang et al. \cite{Jiang} &  & \checkmark &  & Attention & SPECK \\ \hline
         Seok et al. \cite{seok2024novelapproach} &  & \checkmark & \checkmark & ResNet & SIMON, SPECK\\ \hline
         Wang et al. \cite{wang2024investigaingandenhancing} &  & \checkmark & \checkmark & ResNet & SIMON, SPECK \\ \hline
         Maskey et al. \cite{Maskey2025benchmarking} & \checkmark &  &  & GPT, Claude, Gemini, Mistral & AES\\ \hline
         Ours & \checkmark & \checkmark & \checkmark & GPT, Qwen, Mistral  & SPECK \\ \hline
    \end{tabular}
\end{table*}

For RQ1, when we evaluate the performance of the LLM-based neural distinguishers, contrary to expectations, we are unable to observe improvement in the accuracy compared with ResNet in the existing work~\cite{Gohr}. 
We believe that LLMs are primarily for natural language generation and, therefore, might be ineffective for neural distinguishers.
(See Section~\ref{sec:performance-of-LLMs}.)

For RQ2, 
we identify that the choice of differences has little impact on the accuracy of the LLM-based neural distinguisher.
Although the impact of differences on the accuracy has also been discussed in the existing works~\cite{hu2025enhancingneural,lu2025enhanced,wang2024anewrelated-key,wang2024investigaingandenhancing}, our insight is novel because of discovering a new difference that is potentially effective at higher rounds and identifying its impact on LLMs. 
We also confirm that analyzing the distribution of ciphertexts reveals potential for the neural distinguisher to succeed even at higher rounds, depending on the number of data samples.
(See Section~\ref{sec:impact-dataset}.)

For RQ3, we explore prompt designs provided to LLMs and then confirm if the accuracy of LLM-based neural distinguishers is improved by adopting a few‑shot setting~\cite{brown2020language} in which only the XOR operation result between ciphertext pairs is included as an example of differences. 
By contrast, we identify that there are no observable benefits from  Chain‑of‑Thought (CoT)~\cite{wei2022chain}, which guides LLMs through their reasoning step by step. (See Section~\ref{sec:impact-prompts}.)

To sum up, we make the following contributions: 
\begin{itemize}
    \item We propose LLM-based neural distinguishers and conduct extensive experiments. 
    \item For RQ1, LLM-based neural distinguishers provide no observable improvement in the accuracy compared with ResNet.

    \item For RQ2, while the choice of differences has little impact on the accuracy, our analysis of ciphertext distributions suggested that increasing the number of samples has the potential to improve the accuracy.

    \item For RQ3, the accuracy of LLM-based neural distinguishers is significantly improved by a few-shot setting incorporating only the XOR operation result into prompts.
\end{itemize}


\subsection{Related Work}\label{sec:related-work}
Cryptanalysis based on machine learning, including neural distinguishers, has predominantly employed convolutional neural networks (CNNs)~\cite{baksi2022machine,hu2025enhancingneural,kimura2022outputprediction,lu2025enhanced,wang2024anewrelated-key} or Residual Networks (ResNets)~\cite{chen2022anewneuraldistinguisher,hou2025improving,lu2023improvedrelated-key,seok2024novelapproach,wang2024investigaingandenhancing} as architectures.
A wide variety of models have also been explored, such as long short-term memories (LSTMs)~\cite{baksi2022machine,kimura2022outputprediction,pal2024deeplearning-based,watanabe2024ontheeffects}, Bayesian models~\cite{agate2023bayesianmodeling}, LightGBM~\cite{pal2024deeplearning-based}, attention~\cite{Jiang}, and transfer learning~\cite{yadav2025mlbasedimproved,yuan2025rethinking}. 
To the best of our knowledge, the only work that investigates LLMs in cryptanalysis is by Maskey et al.~\cite{Maskey2025benchmarking}, but it is a straightforward evaluation whereby LLMs directly decrypt ciphertexts. 
Namely, it differs from neural distinguishers~\cite{Gohr}, and its accuracy is extremely low, making it quite far from a practical threat posed by LLMs. 
We investigate advanced threats using LLM-based neural distinguishers.

Approaches for enhancing training datasets include the choice of differences~\cite{hu2025enhancingneural,lu2025enhanced,wang2024anewrelated-key,wang2024investigaingandenhancing}, introducing multiple ciphertexts by adding further keys on the same plaintexts ~\cite{baksi2022machine,chen2022anewneuraldistinguisher,lu2023improvedrelated-key}, and enhancing the distribution of samples~\cite{hou2025improving,seok2024novelapproach}. 
We explore differences in accordance with the method by Seok et al.~\cite{seok2024novelapproach}: it applies the principal component analysis and k-means clustering to the dataset in order to select samples. 
We also note that Hou et al.~\cite{hou2025improving} investigate datasets incorporating propagation across different rounds, although we leave it as future work. 
The above discussion is summarized in Table~\ref{tab:related work}.

\section{Preliminaries}\label{sec:preliminaries}
In this section, we provide technical background on LLMs, the SPECK as our target, and neural distinguishers. 
We also describe assumptions of an adversary in this paper. 

\subsection{Large Language Models (LLMs)}\label{sec:llm}
LLMs are natural language processing models that probabilistically generate sequences of tokens based on the frequencies observed in the training data through neural networks.
A token denotes a discrete symbol in a given language and the model generates the next token by selecting ones with the highest conditional probability.
Instructions provided to an LLM are called prompts, and the resulting outputs may vary depending on the prompts.
Representative examples of LLMs include ChatGPT\footnote{\url{https://chatgpt.com}} developed by OpenAI and Gemini\footnote{\url{https://gemini.google.com}} developed by Google.
LLMs have also been used in security research~\cite{chen2024SurveyoflargelanguagemodelsforCyberThreatDetection,naito2023llm-basedattackscenarios,deng2024pentestgpt,shen2025pentestagent}.

\subsection{Short Description of SPECK}\label{sec:speck}
SPECK~\cite{speck-2013} is a lightweight block cipher based on an addition-rotation-XOR (ARX) structure, targeting mainly good performances on microcontrollers.
Following the existing works~\cite{Gohr,benamira2021deeper,hou2021improve,buacuieti2022deep}, we focus on SPECK-32/64, which has a 32-bit block size and a 64-bit key length.
The round function follows a Feistel structure that combines bitwise XOR operations, rotation operations, and 16-bit modular addition.
At round $i$, a 32-bit plaintext is split into two 16-bit words.
The upper 16-bit word at round $i$ is denoted by $l_i$ and the lower 16-bit word by $r_i$, on which the round function is applied.
Subkeys for each round are generated from a master key via a non-linear key schedule.
The main components of this key schedule also employ round functions.

\subsection{Differential Cryptanalysis and Neural Distinguishers}\label{sec:neural}
The process of neural distinguishers is separated into two phases, i.e., difference distinguishing phase and key recovery phase. 
The former is to train a neural distinguisher, and the latter is to employ the trained neural distinguisher to recover the secret key.
We focus on the difference distinguishing phase, and it is formalized as a binary classification problem to distinguish whether, given a ciphertext pair, it has specific differences. 
The objective of the training is to distinguish real pairs, which are ciphertext pairs $(C, C')$ by encrypting plaintext pairs $(P, P')$ that have a difference $P {\oplus} P' = {\Delta}_{in}$ (e.g., ${\Delta}_{in} = 0x0040/0000$), from random pairs, which are ciphertext pairs by encrypting plaintext pairs that do not have the difference ${\Delta}_{in}$.

In the key recovery phase, a new set of ciphertext pairs is generated by encrypting plaintext pairs with input difference ${\Delta}_{in}$ for round $i$.
Candidate keys are then listed and, for each candidate, the corresponding ciphertext pairs are decrypted at a single round. 
The resulting pairs are provided to the trained neural distinguisher to infer whether they originate from plaintext pairs with difference ${\Delta}_{in}$.
Based on this result, the candidate keys are ranked with respect to the similarity between their corresponding ciphertext pairs and the learned differences, and therefore, we can identify the correct key.

\subsection{Assumptions of Adversary}\label{sec:assumption}
The adversary's goal is to build a neural distinguisher with high accuracy.
We assume that an adversary knows an encryption algorithm.
Although a secret key used for encryption is unknown, the adversary can obtain both plaintexts to be encrypted and their resulting ciphertexts. 
In this setting, the adversary does not perform any physical attacks such as side-channel attacks.
The adversary also has sufficient computational resources to train and deploy LLMs.

\section{LLM-Based Neural Distinguishers}\label{sec:llm-basedneural}
In this section, we propose an LLM-based neural distinguisher. 
We first present the design of the LLM-based neural distinguishers.
We then describe the implementation and problem setting.

\subsection{Design of Neural Distinguishers}\label{sec:design}
Our LLM-based neural distinguishers are constructed by fine-tuning a general-purpose LLM with a dataset of ciphertext pairs whose labels represent whether they contain some specific difference or not. 
Since it is necessary to give a system prompt to an LLM, the following prompt is introduced in this paper: 

\begin{itembox}[l]{System prompt}
\textbf{[Instruction]}\\
Please determine if the ciphertext pair comes from plaintexts with a difference of 0x0040/0000 (output 1) or random plaintexts (output 0).\\
Output should be either 0 or 1 only.\\
The encryption algorithm used is 5-round SPECK32/64.\\
\textbf{[Input]}\\
$C$ :0x0051 \textbar~0x35b5 \\
$C'$:0xf417 \textbar~0x64d0 \\
$C$ XOR $C'$ : 0xf446 \textbar~0x5165\\
\textbf{[Output]}\\
Label : '1'
\end{itembox}

In the above prompt, the ciphertexts, i.e., $C$ and $C'$, are provided in the ``[Input]" field.
This field contains both real pairs and random pairs. 
The upper and lower 16-bit words of each ciphertext are separated by `` {\textbar} ", and the XOR results between the upper and lower words of $C$ and $C'$ are included.
In the ``[Output]" field, a value of `1` is assigned if the upper and lower 16-bit words of its underlying plaintext exhibit the specified difference, i.e., ``\textit{0x0040/0000}", and `0` otherwise.
Note that examples provided in the few-shot setting consist of ciphertexts, while the difference to be classified is defined over their underlying plaintexts. Thus, the difference between $C$ and $C'$ may differ from $\textit{0x0040/0000}$ shown at the beginning of the above system prompt.

\subsection{Implementation}\label{sec:implementation}
We used \texttt{mistral-7b}\footnote{\url{https://huggingface.co/mistralai/Mistral-7B-v0.1}}, 
\texttt{mixtral-8x7b}\footnote{\url{https://huggingface.co/mistralai/Mixtral-8x7B-v0.1}}, 
\texttt{qwen3-32b}\footnote{\url{https://huggingface.co/Qwen/Qwen3-32B}}, \texttt{gpt-oss-20b}\footnote{\url{https://huggingface.co/openai/gpt-oss-20b}}, and \texttt{gpt-oss-120b}\footnote{\url{https://huggingface.co/openai/gpt-oss-120b}} as publicly available LLMs.
The number of parameters for each model is as follows: \texttt{mistral-7b} is 7~billion, 
\texttt{mixtral-8x7b} is 47~billion,
\texttt{qwen3-32b} is 32~billion, \texttt{gpt-oss-20b} is 21~billion, and \texttt{gpt-oss-120b} is 117~billion.
We downloaded these models in an AWS EC2 \texttt{p4de.24xlarge} instance as a local environment, which has eight NVIDIA A100 Tensor Core GPUs and a total of 640~GB of GPU memory. 
In this environment, we fine-tune the models using QLoRA~\cite{dettmers2023qlora}.

Following Gohr~\cite{Gohr}, we generated a dataset for our extensive experiments. 
Specifically, on SPECK-32/64 with 5 to 8 rounds, we generate $10^6$ ciphertext pairs for training and $10^5$ ciphertext pairs for testing by taking machine resources into account.
For each round, the dataset is constructed to include equal numbers of real pairs, generated by encrypting plaintext pairs with the difference ${\Delta}_{in} = 0x0040/0000$, and random pairs, generated from plaintext pairs without the difference.
All the models are trained with the following hyperparameters: batch size is 2,560, the number of epochs is 391, learning rate is 0.0001, and optimizer is \texttt{AdamW}. 

\subsection{Problem Definition}\label{sec:problem}

We describe the problem setting below. 
As described in Section~\ref{sec:neural}, we focus on the difference distinguishing phase. 

First, we examine how LLMs affect the performance of neural distinguishers.
In the existing works~\cite{hu2025enhancingneural,lu2025enhanced,wang2024anewrelated-key,wang2024investigaingandenhancing}, the accuracy of neural distinguishers significantly deteriorates after round~8.
By comparing with ResNets~\cite{Gohr} as a baseline, 
we clarify whether the use of LLMs leads to improving the performance of neural distinguishers.
As evaluation metrics, we adopt accuracy and the F1 score.

Second, we examine whether the performance of LLM-based neural distinguishers is improved by enhancing the distribution of a dataset. 
The distribution of ciphertext pairs often depends on the choice of differences and a good dataset exhibits features that can be effectively learned~\cite{seok2024novelapproach}. 
Following Seok et al.~\cite{seok2024novelapproach}, principal component analysis (PCA) and k-means clustering are applied to the ciphertext pairs to identify differences that are effective for difference distinguishing, and then measure accuracy for LLM-based neural distinguishers.

Third, we examine the impact of prompts on LLM-based neural distinguishers.
Prompt engineering can potentially improve the performance of LLM-based neural distinguishers, and we introduce two classic techniques into a system prompt.
The first technique is a few-shot technique, whereby the XOR operation result $C {\oplus} C'$ between ciphertext pairs $(C, C')$ is included in the ``[Input]" field of the prompt. 
The second technique is the Chain-of-Thought (CoT), which instructs a model to reason step by step in the ``[Instruction]" field of the prompt.
We evaluate how these techniques affect the performance of LLM-based neural distinguishers.

\section{Performance of LLM‑Based Neural Distinguishers} \label{sec:performance-of-LLMs}
In this section, we clarify whether the performance of neural distinguishers is improved by LLMs compared with that of conventional neural distinguishers using ResNet. 
We measure the accuracy and the F1 score of LLM-based neural distinguishers using \texttt{mistral-7b}, 
\texttt{mixtral-8x7b}, 
\texttt{qwen3-32b}, \texttt{gpt-oss-20b}, or \texttt{gpt-oss-120b},
and then provide the answer to RQ1.

\begin{figure}[t]
  \centering
  \begin{minipage}{0.93\linewidth}
    \centering
    \includegraphics[width=\linewidth]{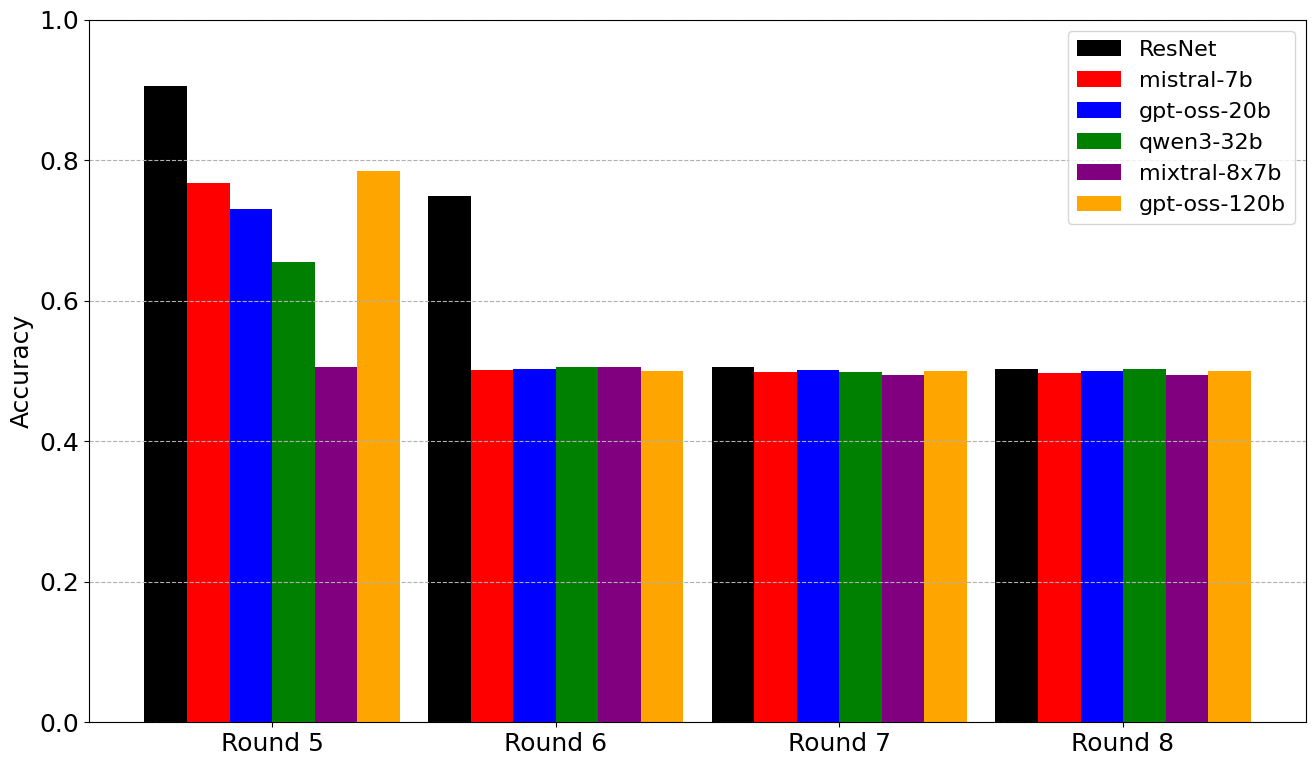}
    \subcaption{Accuracy.}
    \label{fig:accuracy}
  \end{minipage}
  \hspace{5mm}
  \begin{minipage}{0.93\linewidth}
    \centering
    \includegraphics[width=\linewidth]{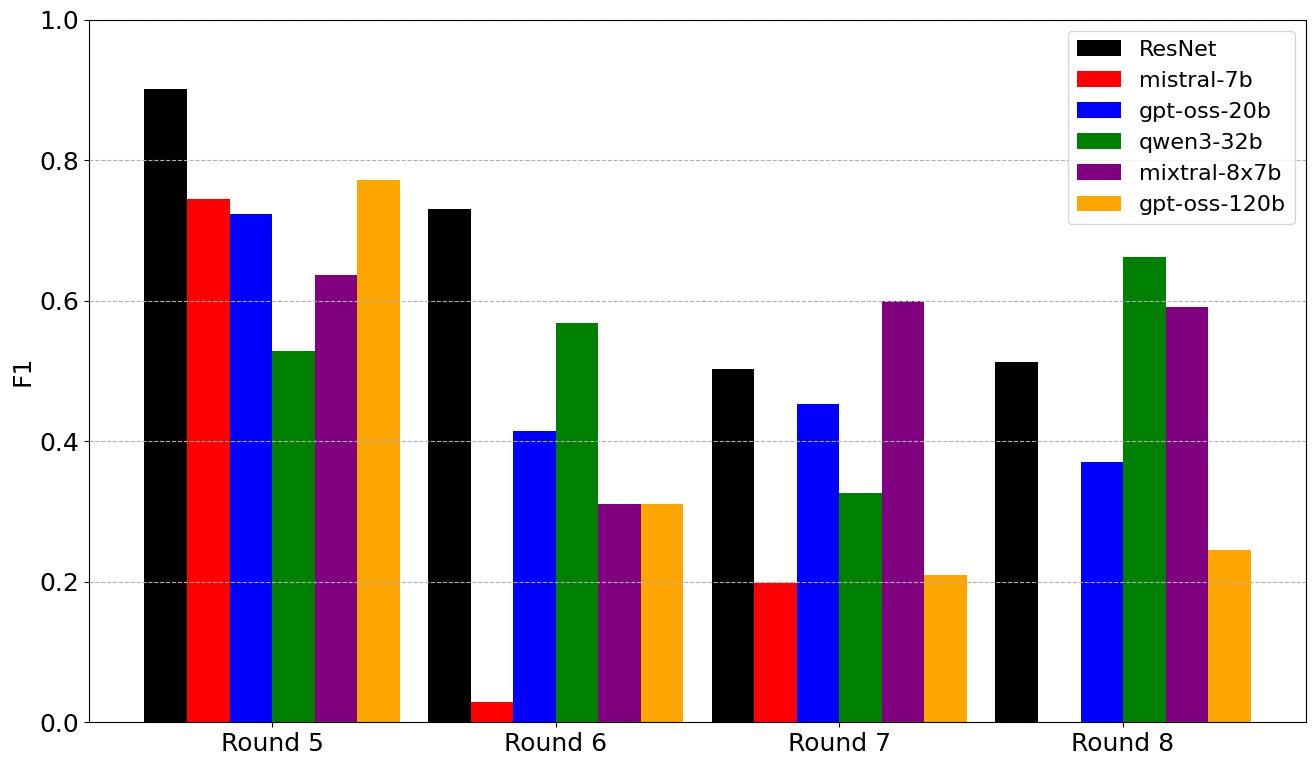}
    \subcaption{F1 score.}
    \label{fig:f1-score}
  \end{minipage}
    \caption{Performance evaluation of each model.}
    \label{fig:performance}
\end{figure}

\begin{table*}[t]
    \centering
        \caption{Candidates of differences for neural distinguishers.
        Each cell represents the differences discovered in each work. 
        Although Seok et al.~\cite{seok2024novelapproach} evaluated many differences in their paper, we picked up only several differences with remarkable scores.}
    \begin{tabular}{c|c|c|c|c}\hline
      Reference & Round 5 & Round 6 & Round 7 & Round 8 \\ \hline
       This work & 0x0040/0000 & 0x0040/0000 & 0x0040/0000 & 0x0000/4000 \\ \hline
      Wang et al.\ \cite{wang2024anewrelated-key} & - & - & - & 0x0040/0000, 0x0040/8000 \\ \hline
      Seok et al.\ \cite{seok2024novelapproach} & 0x0040/0000, 0x0020/0000 & - & - & - \\ \hline
      Wang et al.\ \cite{wang2024investigaingandenhancing} & 0x2800/0010, 0x0211/0a04 & 0x0211/0a04 & 0x0a20/4205, 0x0a60/4205 & - \\ \hline
    \end{tabular}
    \label{tab:input-difference}
\end{table*}

\subsection{Result}\label{sec:results-RQ1}
Figure~\ref{fig:accuracy} and Figure~\ref{fig:f1-score} show accuracy and the F1 score for each model, respectively.
According to the figures, we observe that, except for the F1 scores of \texttt{qwen3-32b} at round~8 and \texttt{mixtral-8x7b} at round~7 and round~8, the performance of the LLM-based neural distinguishers is lower than that of ResNet, which means no observable improvement.
Focusing on the LLMs, we observe that \texttt{gpt-oss-120b} achieves higher accuracy and F1 scores at round~5 than the other models except for ResNet.
It is considered that these results can be obtained by a higher expressiveness of LLMs in accordance with the larger number of model parameters. 
Nevertheless, for round~6 and higher rounds, the other models yield accuracies close to 0.5, which means that ciphertexts are indistinguishable for the neural distinguishers.

Based on our experimental results, we are unable to find evidence that LLMs improve the accuracy and the F1 score of neural distinguishers except for \texttt{qwen3-32b} at Round~8.
Since LLMs are primarily designed for natural language generation, as might have been expected, they sometimes seem to be unsuitable for neural distinguishers.
In addition, several models, including \texttt{gpt-oss} and \texttt{qwen3}, appear to have been pretrained to generate CoT-style reasoning: for instance, LLMs that are pretrained to generate CoT-style rationales may perform unnecessarily complex reasoning even for a simple task such as difference distinguishing, which in the worst case can lead to incorrect inferences~\cite{chen2024not, sui2025stop}.
This observation is also supported by the results: except for \texttt{gpt-oss-120b}, LLMs pretrained for CoT-style reasoning exhibit lower accuracy.

\subsection{Answer to RQ1}\label{sec:AtoRQ1}
According to the above experimental results, contrary to strong performance across a wide range of data analysis tasks, LLMs show no observable improvement in the accuracy and the F1 score of neural distinguishers. 
This is because LLMs are primarily designed for natural language generation and therefore might be unsuitable for neural distinguishers.

\begin{table}[t]
\centering
\caption{Accuracy of each model before and after enhancing the distribution of the dataset.
We show the results on round~5 and round~8.
}
\label{tab:acc_model-datasetadjusted}
\resizebox{\columnwidth}{!}{%
\begin{tabular}{c|c|c|c|c|c}
\hline
\multicolumn{2}{c|}{} &
\multicolumn{2}{c|}{Before enhancement} &
\multicolumn{2}{c}{After enhancement} \\ \hline
Model & Rounds & Accuracy & F1 score & Accuracy & F1 score \\
\hline
ResNet [1] & 5 & 0.906 & 0.901 & 0.906 & 0.901 \\
           & 8 & 0.503 & 0.512 & 0.506 & 0.512 \\ \hline
gpt-oss-20b & 5 & 0.731 & 0.723 & 0.731 & 0.723 \\
            & 8 & 0.500 & 0.371 & 0.500 & 0.487 \\
\hline
\end{tabular}%
}
\end{table}

\section{Impact of Dataset Distribution}\label{sec:impact-dataset}
In this section, as our answer to RQ2, we identify whether the performance of LLM-based neural distinguishers is improved by enhancing the distribution of a training dataset. 
We discuss only ResNet and \texttt{gpt-oss-20b} below.

\subsection{Enhancing Dataset Distribution}
We first execute PCA for the dataset and then execute k-means clustering for the resulting samples of PCA in order to enhance the distribution of a training dataset as described in Section~\ref{sec:problem}. 
The above procedure enables us to select input differences that are effective for neural distinguishers. 
While Seok et al.~\cite{seok2024novelapproach} executed only up to round~5, we executed it on round~6 to round~8. 

As a result, we discovered \textit{0x0000/4000} at round~8 as a new candidate of differences to achieve a higher accuracy. 
It has not been reported in the existing works~\cite{wang2024anewrelated-key,wang2024investigaingandenhancing}, while Seok et al.~\cite{seok2024novelapproach} analyzed the candidates in only round~5.
Table~\ref{tab:input-difference} summarizes the differences discovered.

\subsection{Result}\label{sec:7-1}
Table \ref{tab:acc_model-datasetadjusted} shows the accuracy and the F1 score of the LLM-based neural distinguisher. 
According to the table, we observe that, except for the F1 score of \texttt{gpt-oss-20b} at round~8, enhancing the distribution of the dataset has little impact on the accuracy for LLMs as well as ResNet in the existing work~\cite{wang2024anewrelated-key}.

\begin{figure*}[t]
  \begin{center}
    \begin{tabular}{cccc}
      \begin{minipage}{0.23\hsize}
        \centering
        \includegraphics[keepaspectratio, scale=0.229, angle=0]{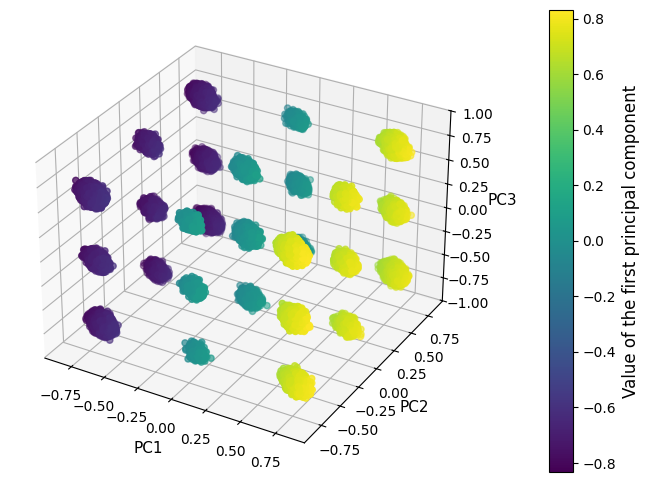}
        \subcaption{Round~5}
        \label{fig:5round-10-5}
      \end{minipage} 
      &
  \begin{minipage}{0.23\hsize}
    \centering
    \includegraphics[keepaspectratio, scale=0.229, angle=0]{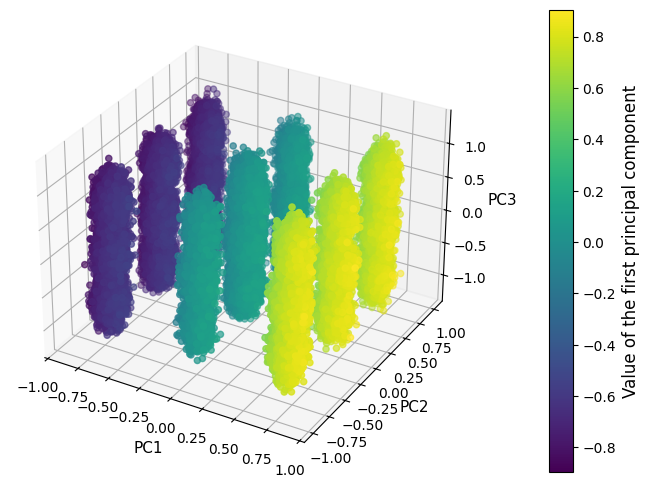}
    \subcaption{Round~6}
    \label{fig:6round-10-5}
  \end{minipage} 
  &
  \begin{minipage}{0.23\hsize}
    \centering
    \includegraphics[keepaspectratio, scale=0.229, angle=0]{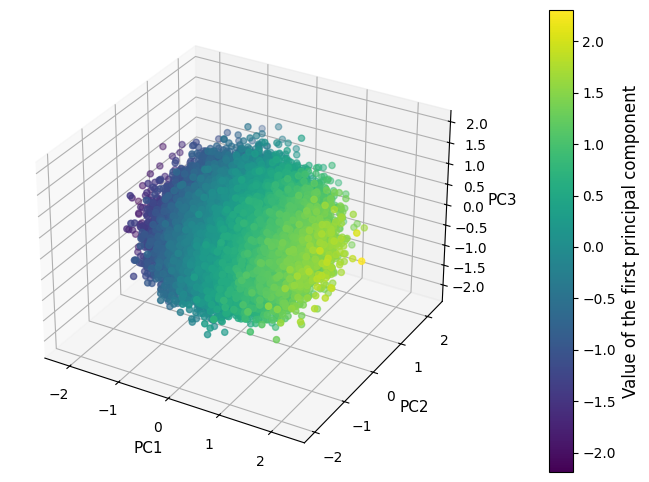}
    \subcaption{Round~7}
    \label{fig:7round-10-5}
  \end{minipage} 
  &
  \begin{minipage}{0.23\hsize}
    \centering
    \includegraphics[keepaspectratio, scale=0.229, angle=0]{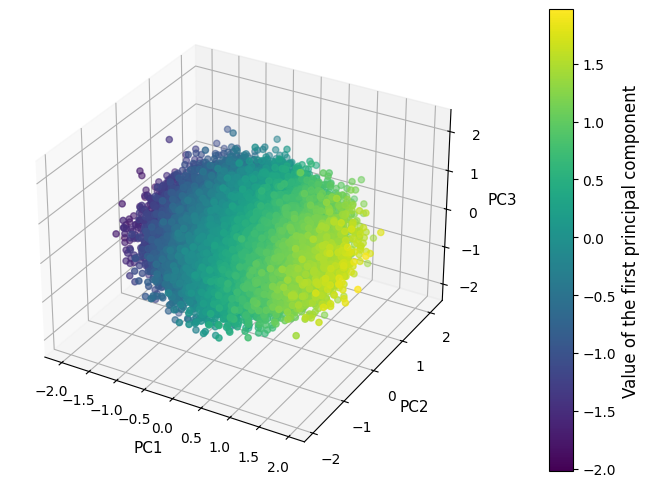}
    \subcaption{Round~8}
    \label{fig:8round-10-5}
  \end{minipage} 
\end{tabular}
\caption{PCA of ciphertexts at each round for $10^5$ samples. }
\label{fig:k-means-10-5}
  \end{center}
\end{figure*}

\begin{figure*}[t]
  \begin{center}
    \begin{tabular}{cccc}
      \begin{minipage}{0.23\hsize}
        \centering
        \includegraphics[keepaspectratio, scale=0.229, angle=0]{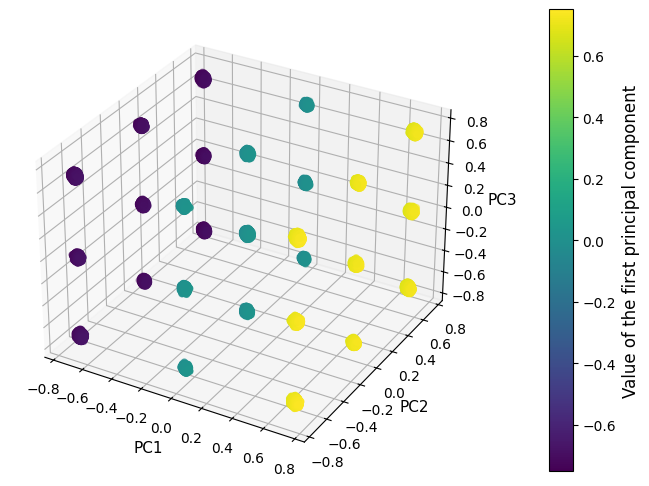}
        \subcaption{Round~5}
        \label{fig:5round-10-6}
      \end{minipage} 
      &
  \begin{minipage}{0.23\hsize}
    \centering
    \includegraphics[keepaspectratio, scale=0.229, angle=0]{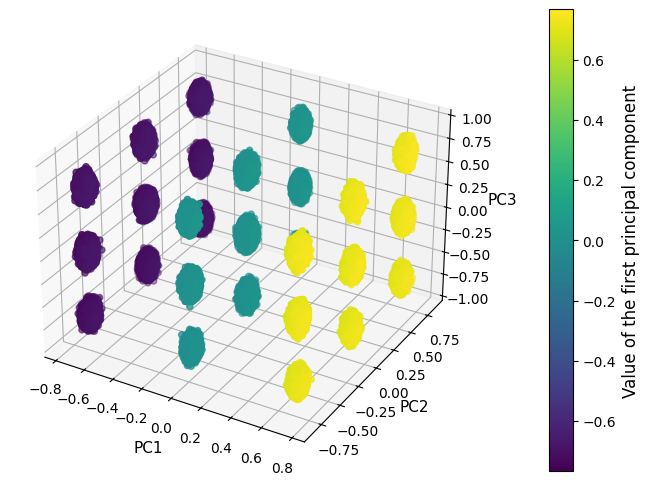}
    \subcaption{Round~6}
    \label{fig:6round-10-6}
  \end{minipage} 
  &
  \begin{minipage}{0.23\hsize}
    \centering
    \includegraphics[keepaspectratio, scale=0.229, angle=0]{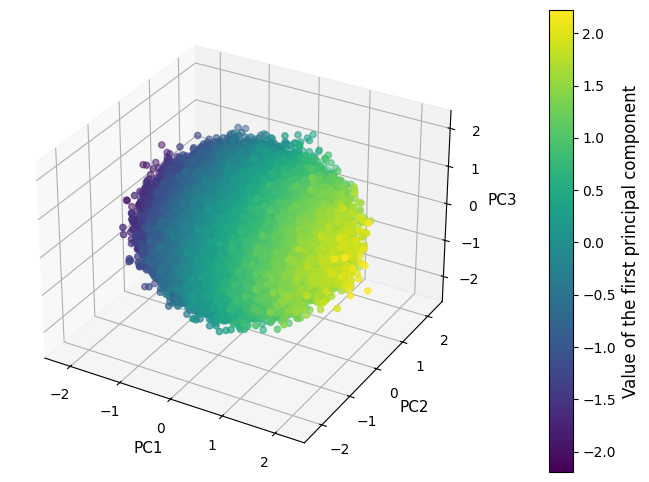}
    \subcaption{Round~7}
    \label{fig:7round-10-6}
  \end{minipage} 
  &
  \begin{minipage}{0.23\hsize}
    \centering
    \includegraphics[keepaspectratio, scale=0.229, angle=0]{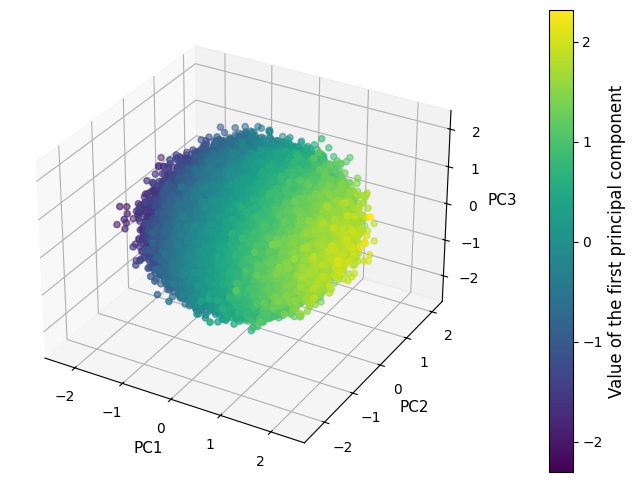}
    \subcaption{Round~8}
    \label{fig:8round-10-6}
  \end{minipage} 
\end{tabular}
\caption{PCA clustering of ciphertexts at each round for $10^6$ samples.}
\label{fig:k-means-10-6}

  \end{center}
\end{figure*}

\begin{figure*}[t]
  \begin{center}
    \begin{tabular}{cccc}
      \begin{minipage}{0.23\hsize}
        \centering
        \includegraphics[keepaspectratio, scale=0.229, angle=0]{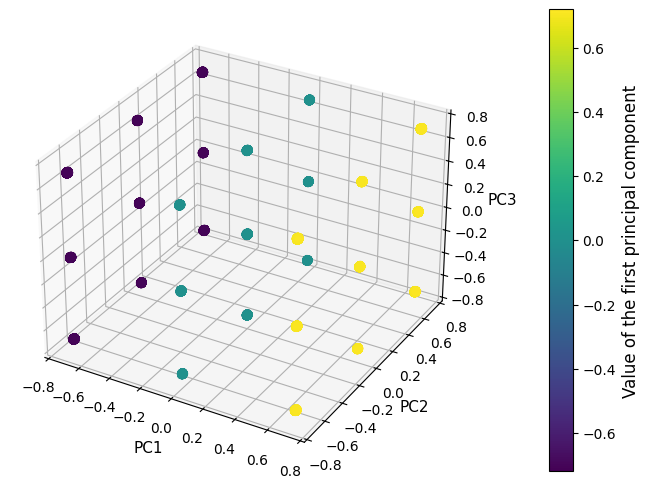}
        \subcaption{Round~5}
        \label{fig:5round-10-7}
      \end{minipage} 
      &
  \begin{minipage}{0.23\hsize}
    \centering
    \includegraphics[keepaspectratio, scale=0.229, angle=0]{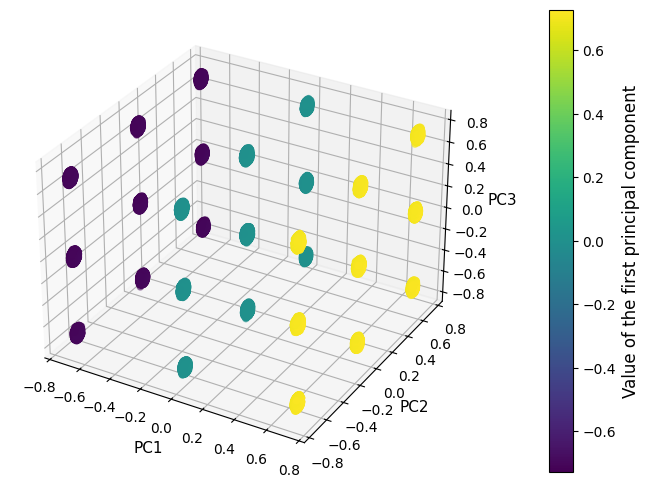}
    \subcaption{Round~6}
    \label{fig:6round-10-7}
  \end{minipage} 
  &
  \begin{minipage}{0.23\hsize}
    \centering
    \includegraphics[keepaspectratio, scale=0.229, angle=0]{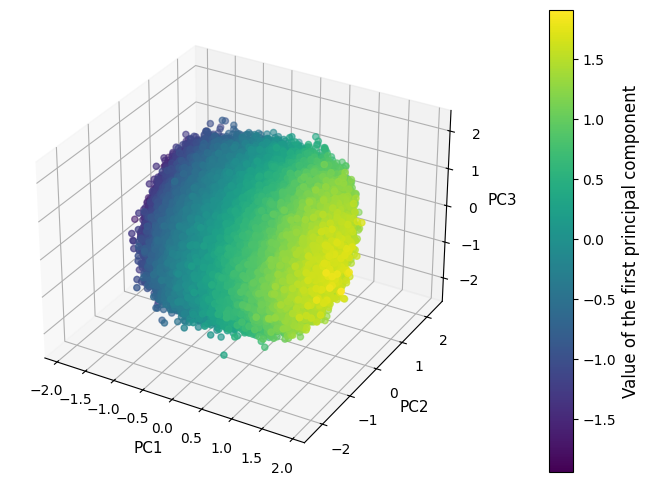}
    \subcaption{Round~7}
    \label{fig:7round-10-7}
  \end{minipage} 
  &
  \begin{minipage}{0.23\hsize}
    \centering
    \includegraphics[keepaspectratio, scale=0.229, angle=0]{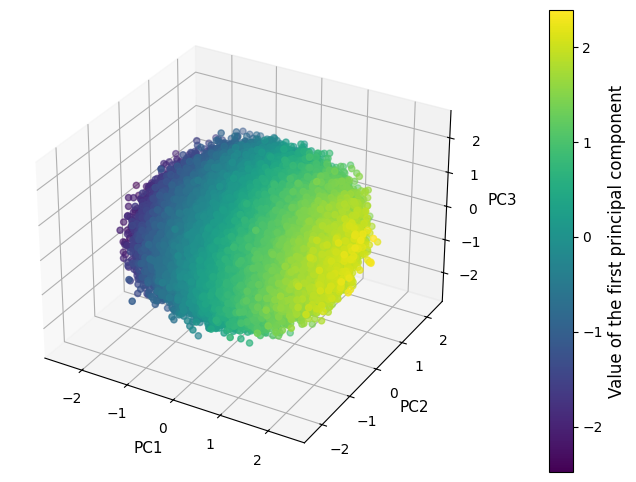}
    \subcaption{Round~8}
    \label{fig:8round-10-7}
  \end{minipage} 
\end{tabular}
\caption{PCA of ciphertexts at each round for $10^7$ samples.}
\label{fig:k-means-10-7}

  \end{center}
\end{figure*}
\begin{figure*}[t]
  \begin{center}
    \begin{tabular}{cccc}
      \begin{minipage}{0.23\hsize}
        \centering
        \includegraphics[keepaspectratio, scale=0.229, angle=0]{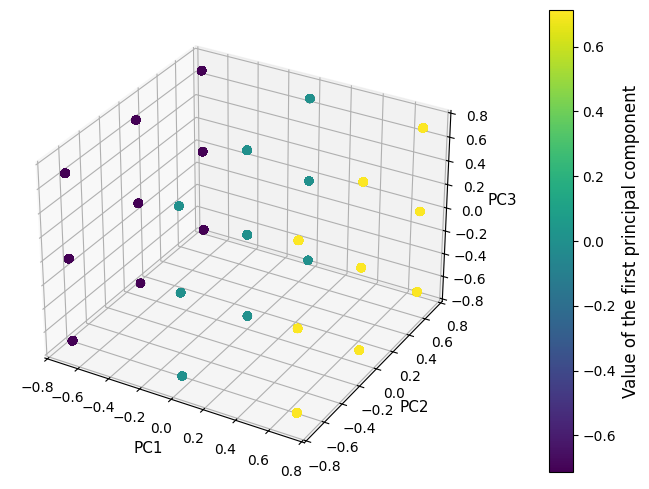}
        \subcaption{Round~5}
        \label{fig:5round-5*10-7}
      \end{minipage} 
      &
  \begin{minipage}{0.23\hsize}
    \centering
    \includegraphics[keepaspectratio, scale=0.229, angle=0]{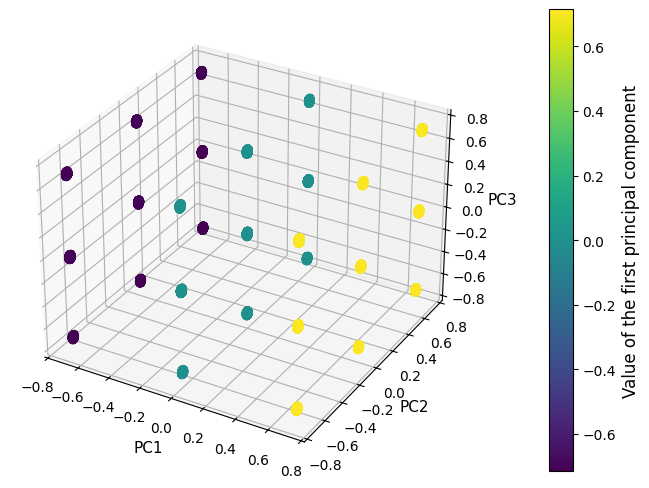}
    \subcaption{Round~6}
    \label{fig:6round-5*10-7}
  \end{minipage} 
  &
  \begin{minipage}{0.23\hsize}
    \centering
    \includegraphics[keepaspectratio, scale=0.229, angle=0]{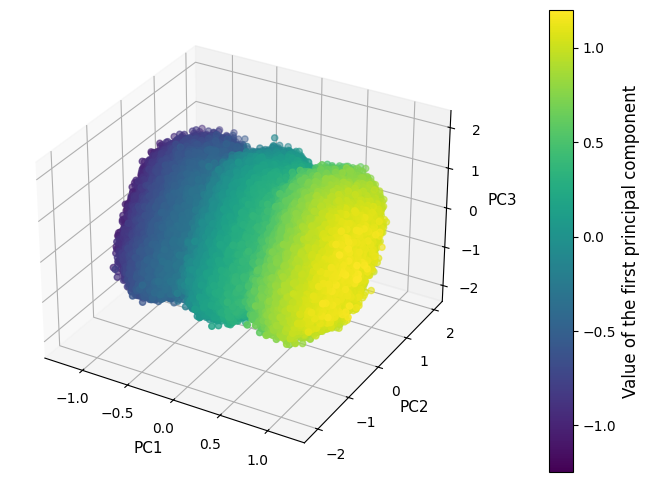}
    \subcaption{Round~7}
    \label{fig:7round-5*10-7}
  \end{minipage} 
  &
  \begin{minipage}{0.23\hsize}
    \centering
    \includegraphics[keepaspectratio, scale=0.229, angle=0]{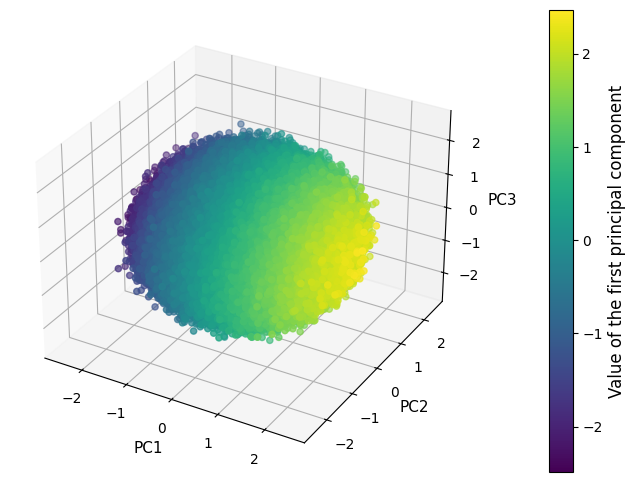}
    \subcaption{Round~8}
    \label{fig:8round-5*10-7}
  \end{minipage} 
\end{tabular}
\caption{PCA of ciphertexts at each round for $5\times10^7$ samples.}
\label{fig:k-means-5*10-7}

  \end{center}
\end{figure*}

To understand this reason, we applied PCA to analyze the distribution of ciphertext pairs for each round. 
Each 64~bit ciphertext pair (as 32~bits $\times$ 2) is projected onto the first three principal components and plotted in 3D space.
Points are colored based on their distance along the first principal component.
Figure~\ref{fig:k-means-10-5} to \ref{fig:k-means-5*10-7} show the results for $10^5$, $10^6$, $10^7$, and $5\times10^7$ ciphertext pairs\footnote{Our computers could only analyze data sets of up to $5 \times 10^7$.}, respectively.
We observe clear distributions in round~5 for the $10^5$ ciphertext pairs and up to round~6 for the $10^6$ ciphertext pairs.
Note that figures indicate that all ciphertext pairs fall within a single cluster from round~7, and the distribution is no longer separated.
However, as the number of samples increases, ciphertext pairs at round~7, which previously appeared random, begin to exhibit emerging structure.

These data distribution analyses imply two observations. 
First, as the number of rounds increases, differences among ciphertext pairs are no longer distinguishable.
This indicates that, regardless of the architectures of neural distinguishers, enhancing the ciphertext dataset at higher rounds is non-trivial within the scope of this paper.
Second, there remains the possibility that increasing the number of samples would enable the difference distinguishing at higher rounds.
For example, we believe that a machine learning model will recognize the distribution of ciphertext pairs even for round~7 and above.
The behavior at higher rounds observed in Figure~\ref{fig:k-means-10-5} and Figure~\ref{fig:k-means-10-6} was not mentioned in the existing work by Seok et al.~\cite{seok2024novelapproach}, and it is a novel part of this paper.

\subsection{Answer to RQ2}\label{sec:AtoRQ2}
Consistent with the existing work~\cite{wang2024investigaingandenhancing}, the choice of differences has little impact on the accuracy of LLM-based neural distinguishers. 
Nevertheless, we identify a new difference that could be effective at higher rounds.
Furthermore, our analysis of ciphertext distributions indicates that, by increasing the number of ciphertext pairs, there remains a possibility that a neural distinguisher can distinguish their distribution even after round~7.

\begin{figure*}[t]
  \centering

  \begin{subfigure}{0.48\linewidth}
    \centering
    \includegraphics[width=\linewidth]{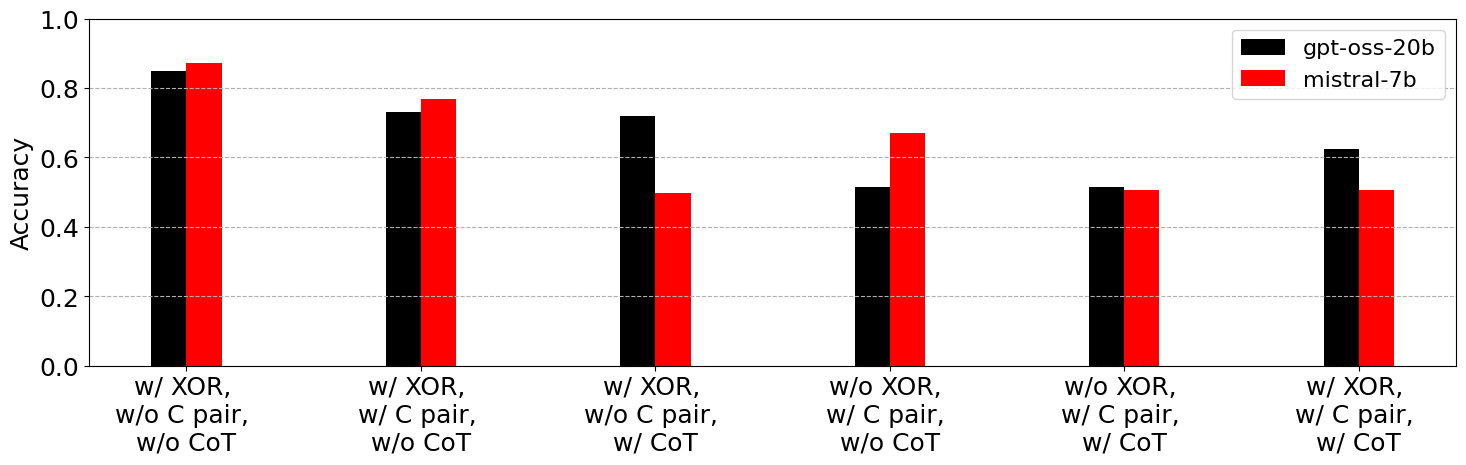}
    \subcaption{Accuracy.}
    \label{fig:acc}
  \end{subfigure}
  \hfill
  \begin{subfigure}{0.48\linewidth}
    \centering
    \includegraphics[width=\linewidth]{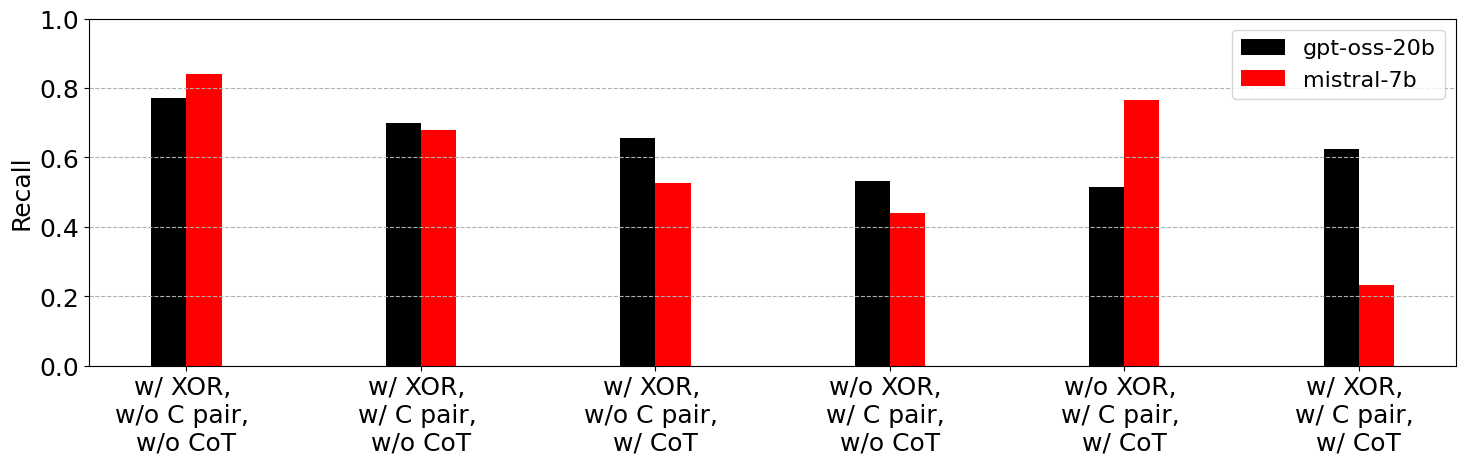}
    \subcaption{Recall.}
    \label{fig:recall}
  \end{subfigure}

  \vspace{2mm}

  \begin{subfigure}{0.48\linewidth}
    \centering
    \includegraphics[width=\linewidth]{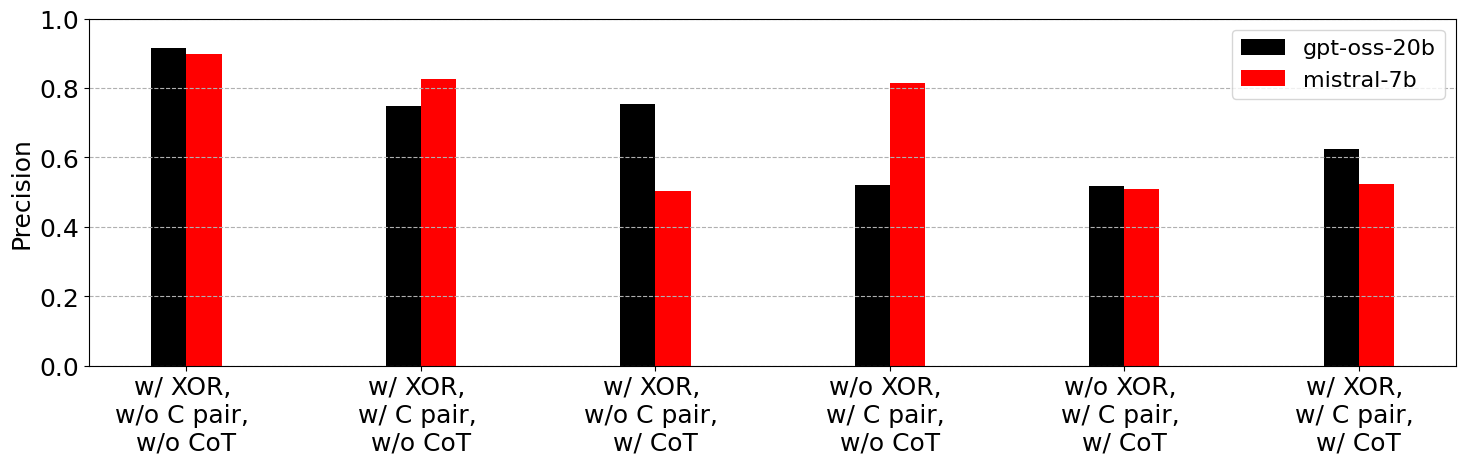}
    \subcaption{Precision.}
    \label{fig:precision}
  \end{subfigure}
  \hfill
  \begin{subfigure}{0.48\linewidth}
    \centering
    \includegraphics[width=\linewidth]{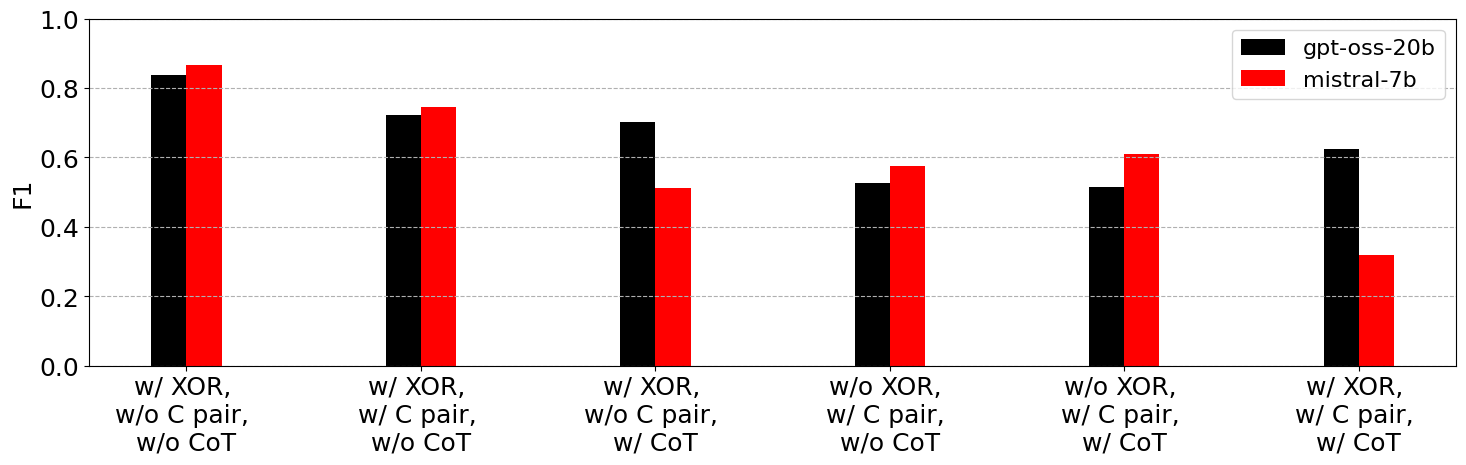}
    \subcaption{F1 score.}
    \label{fig:f1}
  \end{subfigure}
  \caption{The Impact of prompt design on the performance of LLM-based neural distinguishers for round~5.}
  \label{fig:prompt-engineer}
\end{figure*}

\section{Impact of Prompt Design}\label{sec:impact-prompts}

In this section, we investigate the impact of prompt designs on the performance of LLM-based neural distinguishers. 
We focus on \texttt{gpt-oss-20b} and \texttt{mistral-7b} as LLMs on SPECK at round~5. 
We first describe the design of prompts and then present the results for each prompt. 
We finally show the answer to RQ3.

\subsection{Prompt Design} \label{sec:prompt-design}

We adopt two classic techniques for system prompts of LLM-based neural distinguishers, i.e., a few‑shot setting~\cite{brown2020language} and CoT~\cite{wei2022chain}. 
As a few-shot setting, we provide an LLM with the ''[Input]" field in the prompt described in Section~\ref{sec:design}, which consists of ciphertext pairs, the result of the XOR operation between them, or both. 
We note that providing the XOR operation without its underlying real pairs as input data often improves the accuracy of neural distinguishers~\cite{hou2021improve}.
On the other hand, CoT incorporates explanations starting with “Think step by step internally” into the prompt in order to guide the LLM through its reasoning step by step. 
These explanations describe how to compute the XOR of ciphertext pairs in the output, and are split into four steps.

Building on the system prompt described in Section~\ref{sec:design}, we design six prompts by varying the presence or absence of the two techniques above and evaluate their accuracy. 
We do not evaluate prompt settings that exclude the few-shot component because we assume that the difference distinguishing task requires ciphertext pairs and/or differences.
The resulting prompts are provided in ~\ref{sec:appendix}.

\subsection{Result}\label{sec:rq3-results}
Figure~\ref{fig:prompt-engineer} shows the accuracy, the recall, the precision, and the F1 score for each prompt. 
The figure shows that across prompt variants, including the XOR operation results generally leads to higher values for the metrics discussed above, indicating that LLMs base their inferences on the XOR results, except for the \texttt{mistral-7b} settings ``w/ XOR, w/o C pair, w/ CoT" and ``w/ XOR, w/ C pair, w/ CoT".
In addition, we also note that incorporating only the XOR operation results in the prompt as the few-shot setting, i.e., ``w/ XOR, w/o C pair, w/o CoT" in the figure, achieves the highest scores for all the metrics above.
This result is consistent with the existing work\cite{hou2021improve}.

On the other hand, adding CoT tends to degrade performance, except for the recall and F1 score of \texttt{mistral-7b} under ``w/o XOR, w/ C pair, w/ CoT".
We believe that adding CoT makes the instructions redundant, which in turn confuses the LLMs.
However, when comparing the \texttt{mistral-7b} settings ``w/o XOR, w/ C pair, w/o CoT" and ``w/o XOR, w/ C pair, w/ CoT", we observe that adding CoT improves recall and the F1 score.
This may be because adding CoT makes it easier to infer, from ciphertext pairs, the relationship between the output difference and the input difference, suggesting that accuracy could be improved by appropriately designing both the CoT instructions and the information provided.

\subsection{Answer to RQ3}\label{sec:answer-to-rq3}
Our results indicate that adding XOR information in a few-shot setting is effective for prompt design, as all the metrics of the LLM-based neural distinguisher are improved substantially, except for the \texttt{mistral-7b} settings ``w/ XOR, w/o C pair, w/ CoT" and ``w/ XOR, w/ C pair, w/ CoT".
In contrast, even when we added an instruction to compute XOR as part of CoT in the prompt, the accuracy decreased. 
Further investigation of CoT should be undertaken.

\section{Conclusions}\label{sec:conclusion}

In this paper, we investigated whether LLMs pose the threat of neural distinguishers through extensive experiments on SPECK.
We could then obtain the following answers to our RQs.
For RQ1, we observed that the LLM-based neural distinguisher did not improve accuracy compared with the results of ResNet~\cite{Gohr}. 
Next, for RQ2, we confirmed that, at higher rounds, enhancing the distribution of a training dataset has little impact on the accuracy and the F1 score of LLM-based neural distinguishers, although we found a new difference that is potentially effective at round~8. 
We also identified that there still remains a possibility that a neural distinguisher can distinguish differences by increasing the number of samples even after round 7.
Finally, for RQ3, we confirmed that the system prompt incorporating only the XOR operation results as a few-shot setting substantially increased the metrics of the LLM-based neural distinguisher.
In contrast, CoT decreased the accuracy in our experiments. 

In summary, we conclude that LLMs do not pose a threat to machine-learning-based differential cryptanalysis, contrary to expectations. 
We also note that increasing the number of samples may enable neural distinguishers with high accuracy even after round~7, and that refining the design of CoT may be able to improve accuracy; these detailed investigations need to be undertaken in the future.

\textbf{Acknowledgements:} 
A part of this work was supported in part by JST, SAKIGAKE Grant Number JPMJPR23P6 and CREST Grant Number JPMJCR26X7, Japan.

\bibliographystyle{IEEEtran}
\bibliography{reference.bib}

\appendices

\section{Prompt Example}\label{sec:appendix}
We list the system prompts designed in this study.
Here, “XOR” denotes providing $(C {\oplus} C')$, “C pair” denotes providing the ciphertext pair $(C, C')$, and “CoT” denotes adding Chain-of-Thought instructions.

\begin{itembox}[l]{System Prompt (w/ XOR, w/o C pair, w/o CoT)}
\textbf{[Instruction]}\\
Please determine if the ciphertext pair comes from plaintexts with difference of 0x0040/0000 (output 1) or random plaintexts (output 0).\\
Output should be either 0 or 1 only.\\
The encryption algorithm used is 5-round SPECK32/64.\\
\textbf{[Input]}\\
$C$ XOR $C'$ : 0xf446 \textbar~0x5165\\
\textbf{[Output]}\\
Label : '1'
\end{itembox}
\begin{itembox}[l]{System Prompt (w/ XOR, w/o C pair, w/ CoT)}
\textbf{[Instruction]}\\
Please determine if the XOR of ciphertext pair comes from plaintexts with difference 0x0040/0000 (output 1) or random plaintexts (output 0).\\
Output should be either 0 or 1 only.\\
The encryption algorithm used is 5-round SPECK32/64.\\
Think step by step internally:\\
1. Compare the XOR result with the expected difference pattern (0x0040 for left half, 0x0000 for right half).\\
2. If the XOR pattern is consistent with the expected difference (or close to it), output 1, otherwise, output 0.\\
3. Output only the final answer: 0 or 1.\\
\textbf{[Input]}\\
$C$ XOR $C'$ : 0xf446 \textbar~0x5165\\
\textbf{[Output]}\\
Label : '1'
\end{itembox}
\begin{itembox}[l]{System Prompt (w/ XOR, w/ C pair, w/ CoT)}
\textbf{[Instruction]}\\
Please determine if the ciphertext pair comes from plaintexts with difference 0x0040/0000 (output 1) or random plaintexts (output 0).\\
Output should be either 0 or 1 only.\\
The encryption algorithm used is 5-round SPECK32/64.\\
Think step by step internally:\\
1. Compute the XOR of the two ciphertext halves (left and right).\\
2. Compare the XOR result with the expected difference pattern (0x0040 for left half, 0x0000 for right half).\\
3. If the XOR pattern is consistent with the expected difference (or close to it), output 1, otherwise, output 0.\\
4. Output only the final answer: 0 or 1.\\
\textbf{[Input]}\\
$C$ :0x0051 \textbar~0x35b5 \\
$C'$:0xf417 \textbar~0x64d0 \\
$C$ XOR $C'$ : 0xf446 \textbar~0x5165\\
\textbf{[Output]}\\
Label : '1'
\end{itembox}
\begin{itembox}[l]{System Prompt (w/ XOR, w/ C pair, w/o CoT)}
\textbf{[Instruction]}\\
Please determine if the ciphertext pair comes from plaintexts with difference of 0x0040/0000 (output 1) 
or random plaintexts (output 0).\\
Output should be either 0 or 1 only.\\
The encryption algorithm used is 5-round SPECK32/64.\\
\textbf{[Input]}\\
$C$ :0x0051 \textbar~0x35b5 \\
$C'$:0xf417 \textbar~0x64d0 \\
$C$ XOR $C'$ : 0xf446 \textbar~0x5165\\
\textbf{[Output]}\\
Label : '1'
\end{itembox}
\;
\begin{itembox}[l]{System Prompt (w/o XOR, w/ C pair, w/o CoT)}
\textbf{[Instruction]}\\
Please determine if the ciphertext pair comes from plaintexts with difference of 0x0040/0000 (output 1) 
or random plaintexts (output 0).\\
Output should be either 0 or 1 only.\\
The encryption algorithm used is 5-round SPECK32/64.\\
\textbf{[Input]}\\
$C$ :0x0051 \textbar~0x35b5 \\
$C'$:0xf417 \textbar~0x64d0 \\
\textbf{[Output]}\\
Label : '1'
\end{itembox}
\;
\begin{itembox}[l]{System Prompt (w/o XOR, w/ C pair, w/ CoT)}
\textbf{[Instruction]}\\
Please determine if the ciphertext pair comes from plaintexts with difference 0x0040/0000 (output 1) or random plaintexts (output 0).\\
Output should be either 0 or 1 only.\\
The encryption algorithm used is 5-round SPECK32/64.\\
Think step by step internally:\\
1. Compute the XOR of the two ciphertext halves (left and right).\\
2. Compare the XOR result with the expected difference pattern (0x0040 for left half, 0x0000 for right half).\\
3. If the XOR pattern is consistent with the expected difference (or close to it), output 1, otherwise, output 0.\\
4. Output only the final answer: 0 or 1.\\
\textbf{[Input]}\\
$C$ :0x0051 \textbar~0x35b5 \\
$C'$:0xf417 \textbar~0x64d0 \\
\textbf{[Output]}\\
Label : '1'
\end{itembox}

\end{document}